# Non-invasive imaging assisted CFD simulation of 4D multi-modal fluid flow using In-situ adaptor

Vaishali Sharma[1], Arpit Kumar[1], Snehlata Shakya[1], and Mayank Goswami[1, #]

[1]*Divyadrishti Imaging Laboratory,*
*Department of Physics,* Indian Institute of Technology Roorkee,
Roorkee, Uttarakhand-247667, India.

[#]Mayank.goswami@ph.iitr.ac.in

## Abstract

X-ray Computed Tomography (CT) is used to recover the true surfaces of fluid channels and fed to simulation tool (ANSYS) to create accurate cyber environment. The simulation tool also receives CT-assisted multiphase fluid profiles (belonging to the instance just before the flow starts) as an initial condition.

This unique methodology is made possible by using a novel in-situ compact adaptor design is used to create fluid channels that can be placed inside any industrial X-ray CT and fulfill the above objective. It is integrated with an android based App to control the flow once placed inside CT. It is portable and compact enough: (a) to be placed inside various experimental environments, and (b) modular enough to be mounted with multi-modal systems simultaneously.

Two key parameters: (a) spatial distribution and (b) the air volume fraction, are measured using two different non-invasive imaging modalities: (a) Electrical Impedance Tomography (EIT) and (d) X-ray Computed Tomography (CT). Simulated outcomes are correlated with the experimental outcomes from both EIT and X-ray CT, showing an agreement of 85 to 98 percent, respectively. Time-averaged electrically conductive fluid flow profile obtained by EIT shows a match with mass attenuated fluid profile obtained by X-ray CT, justifying the utility of an in-situ adaptor. CT assistance for CFD studies can be replaced by EIT assistance as former techniques: (a) scanning time may be relatively slower than the latter, (b) it does not require rotations, (c) economical, and (d) fluid channels need not to be placed inside of shielded compartment thus improving practicality. The data of analysis is shared in this work.

Multimodal non-invasive imaging provides multiphase flow information, it also differentiates conductive, and mass-attenuated multiphase profiles at common cross-sections.

## 1. Introduction

Multiphase fluid flow comprises the movement of a variety of phases (e.g., liquids, gases, solids, all coexisting phases) [1] within a single pipeline. In various industries, especially those involving extensive and complicated pipelines[2,3] multiphase flow is a very common occurrence[4,5] there. Using the same system/pipeline, the mixture of all phases, like solids, liquids, and gases, are being transported. There are numerous key aspects and challenges [6] associated with this process, such as spatial distribution of phases [7,8], interaction between different phases [9,10], identifying flow patterns [11,12], pressure drop[13,14], safety concerns, and many more. The study of these aspects is important to understand the behavior of the fluid flow, process optimization, quality control, safety purposes, process monitoring, and other factors. This study becomes more complex if the chamber (pipeline) size is large, and inspection of these pipelines is not easy. The complexity of the pipeline arrangements, structures, and material present inside makes it impractical to open them up for inspections [15]. The major problem occurs when these pipelines face issues, such as blockages or disturbances in the flow. Impurities, pressure changes, foreign materials, or some other factors might be the reason for the same. In those cases, it becomes crucial to understand how the different phases within these pipelines flow and interact.

The study of the multiphase flow with a complex fluid chamber is challenging [16,17]. The equipment for real-time measurement may not always be handy or easy to mount on to the fluid flow channel. In the worst-case scenario, operational conditions may destroy the equipment. For example, imaging systems mounted on primary or secondary channels inside radiation-prone locations of nuclear reactors. In such cases simulation may help to predict valuable design and control information. However, the accuracy of the simulation depends on several factors and requires correlations such as slip value between phases. The experimental setup in a laboratory environment is used to estimate accurate correlations [18]. Limitations of simulation and experimental techniques are briefly discussed below:

## 1.1.  Computational Fluid Dynamics

The choice for the size of the mesh, boundary conditions (initial velocity, viscosity, temperature, and other physical parameters) matching with real life, and selection of appropriate model are a few important factors affecting the accuracy of any computational fluid dynamics (CFD) simulations. Accurate representation of the boundaries, along with the proper mesh refinement (according to the shape and size of the object), is a challenging task in most mesh generation methods[19,20]. Inaccurate details of the parameters and models related to meshing can cause numerical divergence and weak correlation with reality. Optimal and autonomous mesh generation covering the channel is crucial due to the potential increase in computational cost associated with complex meshing [21]. Most of the time, correlations between imaging results and simulation results are performed by independent authors [22,23]. Structural irregularities, especially the interior wall of the multi-phase channel, are not accounted for in the simulation tool to create a relevant mesh in the cyber world. The inner wall of a channel may not be as regular as the outer wall due to possible corrosion, accumulation of algae or other substances, damage, etc. Such irregularity may affect the natural circulation, thus affecting the multi-phase flow and its distribution. If not included in the simulation, this may result in inaccuracy [24] . In operator-written codes, these conditions are added, e.g. setup geometry. However, transient conditions close to the real world may not be set. In commercial simulation tools, flow related initial conditions are usually set by choosing options from the in-built drop-down menu. At best, a sparger design is utilized to create a transient flow that leads to steady-state analysis. Limitations of mesh-free CFD simulations such as requirement of global, or local background mesh for integration, lower modelling accuracy and difficulties in boundary representations are discussed in a recent work [25].

## 1.2.  Experimental Techniques

### 1.2.1.  *Invasive methods*

Laser-based interferometry, Schlieren Imaging, and Particle Image Velocimetry (PIV) are the existing supportive modalities used for fluid flow imaging [26]. To perform Laser interferometry, the fluid must be seeded with particles, which may change the fluid's inherent behavior [27]. Schlieren imaging entails modifying the flow by adding deflectors or a knife edge [28]. Tracer particles are required for PIV, which could change the flow properties [29]. There may be difficulties preserving the integrity of the observed fluid dynamics due to the invasive character of these methods.

### 1.2.2. NDT & CT

Alternatively, several non-invasive techniques are available. The spatial distribution [26,27] of the phases, along with the air volume fraction [30,31] inside the tube, can be measured in real time. Ultrasound non-destructive technique (NDT) is deployed for real-time applications. It, although, needs a continuous layer of coupling with the surface of the channel and results in 2D qualitative and/or quantitative information at best [32]. In some rare cases, ultrasound, Gamma-ray, or X-ray computed tomography techniques are also employed to create 3D flow distribution, but these techniques may suffer from an acquisition speed-related dynamic bias issue [33]. Ultrasound, Gamma, and X-ray provide acoustic impedance and energy-dependent mass attenuation distribution of fluid. Gamma-ray CT has limited spatial resolution. X-ray CT enables the visualization of the dynamics of the fluid flow with the capability of both static and dynamic imaging, offering relatively better depth of penetration. It is used to study the multiphase flow and annular flow patterns and measures the fluid velocity along with assessing the condition of the fluid [34]. The scanning speed of a typical X-ray CT, however, remains relatively slower; thus, it is only useful to study relatively slower flow.

The Electrical Impedance Tomography (EIT) provides electrical impedance distribution properties of the material [35]. The EIT system does not need rotating objects because electrodes are already distributed over the periphery to cover the object. It is one of the fastest (micro-second per scan), cheapest, and lightest non-invasive imaging techniques that have a wide range of applications [36], especially in the study of multiphase flow, but with limitations in spatial resolution [37]. So far, it has been used to study slurry bubble columns [38], where it shows the interaction between liquid and bubbles, passive cyclonic gas-liquid separators [38], mapping of different phases, blockage in the tube, and others.

## 1.3. Laboratory setups/adaptors and multi-modal NDT

Researchers have established bulky fluid flow chambers and studied the behavior of fluid inside those channels/chambers [39,40]. Most of these chambers are made of stainless steel and range in length from 7 to 22 meters with intricate pipeline systems. The examination of these chambers presents a few shortcomings. Firstly, large chambers are difficult to handle due to their sheer size, weight, and cost; limiting their practicality, especially in R&D environments where space is a constraint. Secondly, the slow, time-consuming, and laborious operation hinders the effectiveness of the research. Operating these chambers may pose a safety concern if not managed or maintained properly. The ionizing modality with the strong radiation source may pose additional safety issues requiring bulky shielding. Therefore, in a laboratory context, a small design that can mimic and replace the intricate channels is required to study the behavior of fluids. A relatively small chamber design is suggested, with both gamma-ray tomography and EIT system mounted along the channel length at different locations, giving the inner profile of different cross-sections. It may be useful for quantitative verification from sparger till the point of measurement only but not at the very point of measurement [41]. It also requires multiple gamma-ray sources posing radiation hazards in surrounding safety protocol to house and handle. Laboratory setups mimicking the industrial channels exactly in shape, size, and weight, and thus, necessary jacket/chamber type NDT/CT system would be costly, bulky, pose radiation hazard, impractical to handle, and may require a lot of laboratory space. Scanning speed may also be slower if the scanning array rotates around the channel for a single scan.

## 1.4. Motivation

A corroded interior wall of the channel, dented during the operational period, or having naturally existing irregularities, if not included in the simulation model, may provide an inaccurate flow simulation. Designing such an internal surface may not be easy, even for CAD experts.

A small compact, in-situ multiphase flow chamber can provide a controlled, easily accessible, and scaled-down replicable experimental environment for studying numerous multiphase flow scenarios and assisting simulation tools for improved correlation. The adaptor, thus, is utilized to present X-ray CT-assisted simulation. The 3D printable scaled-down replicas can be created by X-ray CT imaged of the real fluid flow channels.

The secondary motivation of this study is to test a modular compact in-situ adaptor for multi-phase flow imaging using an EIT belt when it is housed inside an X-ray CT to scan a common cross-section and present a true comparison of same flow conditions. These two motivations are the novelty of this work.

## 2. Material and Methods

Details of in-situ adaptor, simulation, and imaging are given in this section. Experiments using developed in-situ adaptors are performed, and simulation is done using commercially available ANSYS software to improve this correlation. This solution addresses the issues related to the installation of the fluid chamber, replication of the simulated results with accuracy in the real world, and user-friendly, easy control of the chamber, especially for research purposes. The user-friendly Android-based application is used to control the device. Its simplifying involvement makes the study of fluid chambers more straightforward. Two key aspects, spatial distribution and the air volume fraction, are simulated and compared with the experimental outcomes using X-ray CT and EIT belt.

## 2.1. Experimental Setup

### 2.1.1. Development of in-situ adaptor:

The fluid chamber mainly has three parts: (1) a cylindrical chamber that contains a tube attached to a pump, (2) a control unit, and (3) a smartphone-based Android app to control the fluid chamber.

#### 2.1.1.1. In-situ adaptor design:

The real-world image of the adaptor, along with CAD, is shown in Fig. 1(a). It is divided into two parts: supportive structures for mounting the chamber and on its top mount controlling unit. The chamber is a cylindrical hollow structure. The hollow structure of the box can be filled with any fluid or amorphous material, such as slurry/sand/earth/mud, depending upon the available energy of the X-ray source or excitation voltage and injecting current of EIT electrodes. For our study, it is simply filled with saline solution. SolidWorks software is used to create Computer-Aided Design (CAD) files for the base Fig. 1(b). The bottom of the supportive structure has a diameter of 7 cm, the middle part of 9 cm, and the cylindrical hollow structure has a diameter of 5 cm. These dimensions, however, can be modified as per the chamber size of X-ray CT and/or length of the belt of EIT. Ultimaker Extended 2+ is used to print three-dimensional structures. The parameters for layer height, fill density, and fill pattern are kept at 150 mm, 45 mm/s, and 80 percent, respectively. The bed temperature and nozzle temperature are set at 45°C and 235°C, respectively. Ploy-Lactic Acid (PLA) ink is used to print the samples, and it has a 0.2 mm nozzle.

The cylindrical chamber is a plastic cylindrical shaped bottle with a diameter of 4.8 cm and a wide opening cover. The cover has two holes to fix the silicon tube inside. The diameter of each hole is 1cm. This box is glued to the hollow structure. A silicon tube with a diameter of 0.5 cm is fixed inside the cylindrical box.

#### 2.1.1.2. Control unit:

The control unit of the fluid chamber contains (a) a microcontroller to execute the controlling codes, (b) H-bridge circuit, (c) Li-ion batteries for powering the pump, (d) charging module to charge the batteries, (f) Bluetooth module to transfer controlling instruction in real-time when the in-situ adaptor is inside the X-ray CT chamber while the X-ray source is on, and (e) peristaltic pump to induce the flow of fluid.

To hold all electronic components a CAD model of the base has been prepared. This base is basically a rectangular box having length, width, and height of 10 cm, 4 cm, and 2.5 cm, respectively but it depends on the size of the adaptor. It contains two holes with a diameter of 0.5 cm from the center to the side to provide a gateway for the silicon tube and is placed right above the cylindrical chamber. Ultimaker Extended 2+ is used to print three-dimensional structures with the same specifications described in the previous section with same PLA material. Bluetooth module contains 6 pins, among which 4 pins (Rx, Tx, GND, and Vcc) are connected with Arduino. Vcc, GND, Rx, and Tx are connected to 5v, GND, 9, and 10 pin of Arduino, respectively. H-bridge is also connected with Arduino. Pin ENA, IN1, and IN2 of the H-bridge are connected to 5, 3, and 2 pin of

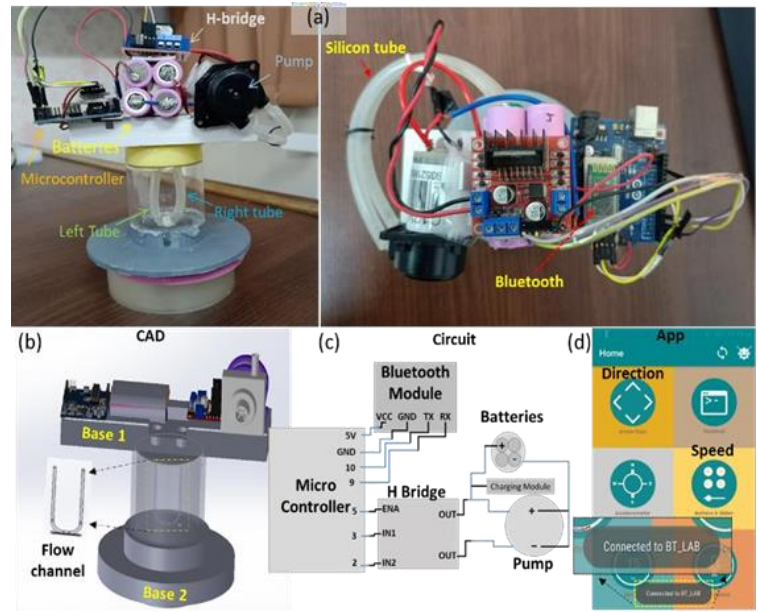

**Figure 1**: Single channel in-situ adaptor: (a) Real photo, (b) CAD, (c) Circuit, and (d) Screenshot of controlling app.

Arduino, respectively. The output of the H-bridge (OUT2) is connected to both the terminals of the pump's input. Four Li-ion batteries are connected in series, and the output of the batteries is also connected to the input of the pump. All four batteries are connected to the charging module to recharge them. The output of the pump is connected to a silicon tube. A code (weblink) is written and uploaded in Arduino IDE to control the assembly. The complete assembly of the control unit is shown in CAD model in Fig. 1(b). The block diagram of the circuit is shown in Fig. 1(c).

*2.1.1.3. App-based control:*

The control system relies on an Android-based user-friendly app that controls the speed and polarity of pump thus velocity and direction of fluid inside the chamber. The control signals are ascii values, each of them gets interpreted as a particular action (e.g. start, stop, increase/decrease speed). The App establishes a wireless connection with adaptor providing the control. Direction control has forward and reverse direction modes, while there are four control modes for speed. Screenshot of this app while connecting with in-situ adaptor (named BT_LAB) is shown in Fig. 1(d). An open-source Arduino Bluetooth control platform is used to develop this interface [42].

## 2.1.2. EIT:

The EIT setup is designed with 16 aluminum electrodes having dimensions of 1cm in length and 0.5cm in width. For the experiment, EIT electrodes are fixed at the inner periphery of the cylinder of the fluid chamber (electrodes can be distributed at the outer periphery of the cylinder as well, and the results in both cases do not differ significantly), and the chamber is filled with the saline solution. Electrodes are subjected to an injection of a constant AC current of ~1mA at a frequency of 50 kHz and voltage $4 V RMS$. The adjacent electrodes strategy [43] to excite the electrodes and data collection is used. Data is collected using an in-house developed switching module. It takes only a few seconds (10 seconds) to complete one set of measurements. This duration can be further reduced in microseconds provided the available budget to improve the electronics. The estimated spatial resolution of this EIT setup is 2 mm.

### 2.1.3. X-ray CT:

The X-ray $\mu$CT is integrated with a solid-state detector (Teledyne RadEye 1 (EV)) to detect radiation generated by 35kV voltage and 1 mA tube setting. The CT measurements are performed by rotating the fluid chamber to 360-degree. The number of rotations is set to perform this measurement under 2 minutes while maintaining 2 mm spatial resolution capability in reconstructed images having 1000 x 1000-pixel resolution. Experimental arrangement of X-ray CT is shown in the result section.

| Table 1 | |
|---|---|
| **Parameter** | **Assigned Value** |
| Physics preference | CFD |
| Mesh details | Solver preference: Fluent Solver: Pressure based Growth rate: default Smoothing: medium Method: CutCell |
| Velocity formulation | Absolute |
| Initialization method | Custom/Hybrid |
| Total time | 600 seconds |
| Time step size | 0.001 |
| Gravity | Included (z-axis) |
| Phases | Sanitizer (primary) Air (secondary) |
| Flow rate | 2 mm/2 min or 0.03 mm/sec |

## 2.2. Simulation details:

### 2.2.1. Meshing of ROI

The empty and filled in-situ adaptor is placed inside the X-ray CT machine, and data is collected and reconstructed. Images are obtained in 2D slices. A segmentation code is written, and the tube's inner wall, outer wall, and fluid profile (more details are given in section 2.4) are segmented. The segmented 2D slice binary images are loaded in Fiji software and saved as the STL file. The STL file is not usually well-defined due to irregular segmentation showing misaligned rough surface. Due to this STL files cannot be handled using any CAD software. An open-access software, MeshLab, is used to remove the surface roughness. The STL file is then uploaded in SolidWorks software using "scan to 3d". A well-defined regular shaped tube is created and imported into ANSYS software environment. The simulation geometry is defined through (perpendicular) the bubble plane so that after simulation, a cross-section can easily be seen. Meshing is done afterwards using the CutCell method (the default meshing method of the ANSYS). The flow mechanism is set as per according to pump having higher pressure at one end and lower at another end. Information of the phases, density, and viscosity of the fluid is provided in the next section. Details of other necessary parameters are given in Table 1. The filled in-situ adaptor is scanned, and initial fluid profile is segmented from the tube. This fluid information is fed into the simulation tool as transient initial condition.

## 2.3. Two phase Fluid

A gel-based hand sanitizer of 95% (v/v) of ethyl alcohol (viscosity 1.074 mPa.s at 25 °C) is used. A slight amount of air is introduced (allowed to leak into) in it. This air-in-gel mixture forms a viscous enough fluid so when it moves, bubbles in motion retain their structure and size relatively longer with respect to time, unlike a bubble in a less viscous fluid such as water that has a higher probability of collapsing to nearby bubbles in relatively swiftly manner. So, this mixture offers chance of getting imaged in three conditions: bubbles (a) moving in fluid without collapsing, (b) collapsing after certain time if speed is changed, and (c) sustaining the post collapsing condition; a good characteristic to study while imaging [44]. The flow is basically laminar with high viscosity as reversing the direction of pump results in almost the same structure for low speed movements.

## 2.4. Fluid flow imaging:

Approximately 8 ml of fluid is filled inside the channel. Air bubbles are allowed to percolate inside the tube for two days and the entry point is sealed. Gravity assisted uniform bubble distribution is acquired inside channel. Control code is written and uploaded in a microcontroller using Arduino IDE. The android-based application is launched and connected to adaptor's pump via Bluetooth.

The adaptor pump is started clockwise for 10 seconds assuming it will stir the fluid-air profile causing some of the bubbles to collapse and create a more realistic non-uniform 3D profile. Afterwards, the pump is kept off for 10 minutes. EIT belts are mounted on the adaptor and placed inside the X-ray CT. The pump is re-started, this time in counterclockwise direction, for next 120 seconds and then stopped. Meanwhile multi-modal scanning is performed. The locations of EIT belt mounted on chamber periphery is shown in Fig. 2. It is assumed that peristaltic pump results in steady flow.

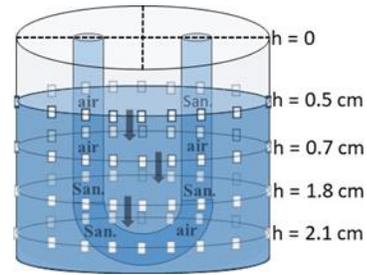

Figure2: EIT electrode distribution on fluid chamber.

# 3.    Results and discussion:

## 3.1.    Imaging assisted simulation

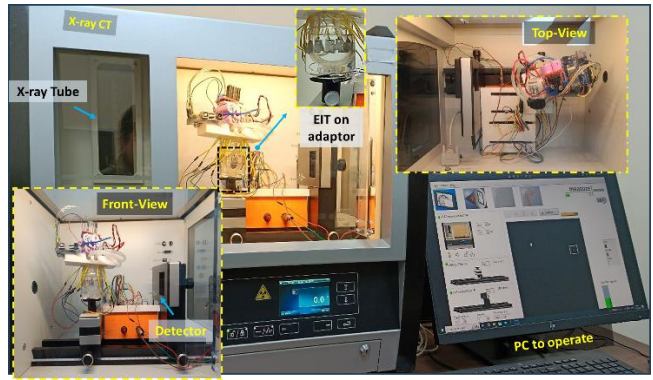

The first stage of experiment is performed by mounting the EIT system on in-situ adaptor and placing it inside X-ray CT, simultaneously as shown in Fig. 3. Once the multi-modality measurements are finished, X-ray data is used to create transient / initial conditions and simulate the flow. The movieA shows the experiment (https://youtu.be/mPuNTeRf2HM). Results of the simulation (air bubble flow), along with the X-ray $\mu$CT results, are shown in Fig. 4(a). The entire time series in video format is shown in movieB. The first row shows the real-world image (through a DSLR camera) of the fluid tube, and the second row shows the X-ray CT images. Third row shows the presence and flow of the air bubbles (simulated) for the same physical situation (column wise).

Figure 3: Experimental Setup (EIT setup inside X-ray CT machine for simultaneous measurements)

Simulated results show good agreement with the experimental results up to 2 minutes. After 2 minutes, X-ray CT and real world images show match while simulation tools integrated/ inbuilt model starts showing disagreements. The 2D slices of simulated images, X-ray CT and EIT images on different heights (marked by black pointers in top row of Fig. 4(a)) are shown in Fig. 4(b). The first row shows the simulated cross-sectional view, second row shows the average cross section of X-ray CT images, and the third row shows corresponding EIT images. The red, dark black, and light blue fake color scheme represents the presence of the air in simulated, X-ray CT and EIT images, respectively. The attenuated profile of the sanitizer is represented by blue, gray, and orange bright spots. The tube region is segmented. To differentiate between air and sanitizer, thresholds are obtained by scanning an empty tube (air) and a fully filled tube (sanitizer) using both X-ray and EIT separately. For X-ray images, pixels with values less than 4000, while for EIT images, pixels with values less than 0.5 are classified as air. These thresholds help identify air-filled spaces in the images accurately. For the calculation of air-volume fraction, the left and right parts of the tube in the image are selected based on the change in pixel values from the background. The fluid part is segmented from the tube. Then, the total number of pixels in the selected region (fluid on both the right and left sides) is separately counted. Pixels are labeled as air or sanitizer based on their values relative to the thresholds.

Subsequently, the number of pixels classified as air and sanitizer are separately counted for each side. Finally, the total number of air pixels is divided by the total number of pixels in the corresponding side to determine the proportion of air pixels.

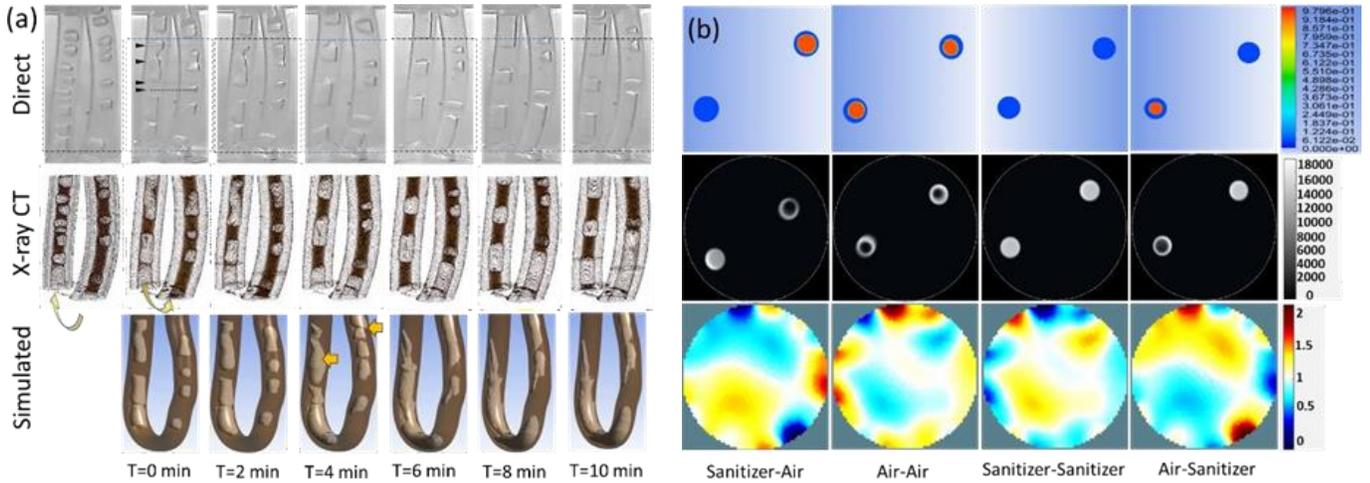

Figure 4: (a) 3D fluid flow Imaging: (i) top row: real-world images, (ii) middle row: 3D X-ray CT images, and (iii) bottom row: Simulated (air bubble flow) images, (b) 2D slices of fluid flow imaging: (i) top row: simulated cross-sectional view, (ii) middle row: mass attenuation distribution by X-ray CT, and (iii) bottom row: corresponding conductive and non-conductive phase distribution in EIT images.

The calculated air volume fractions are shown in Fig. 5(a). It shows that the simulated air-volume fraction of a 3D volume shows 85 to 98 percent agreement with the experimental results despite significant differences in 2D flow profiles, highlighting the need for cross-sectional imaging for accurate analysis. The mentioned agreement increased with an increase in EIT electrode size. The air-volume fraction with respect to radial distance is also plotted in Figures 5(b) and 5(c). The figure shows the because of its good penetration depth, X-ray CT has good agreement with simulated outcomes, whereas EIT has its limitations in terms of penetration. However, even though the spatial resolution of EIT is inferior to the performance considering flow rate vs accuracy to estimate air fraction is comparable with slow X-ray CT.

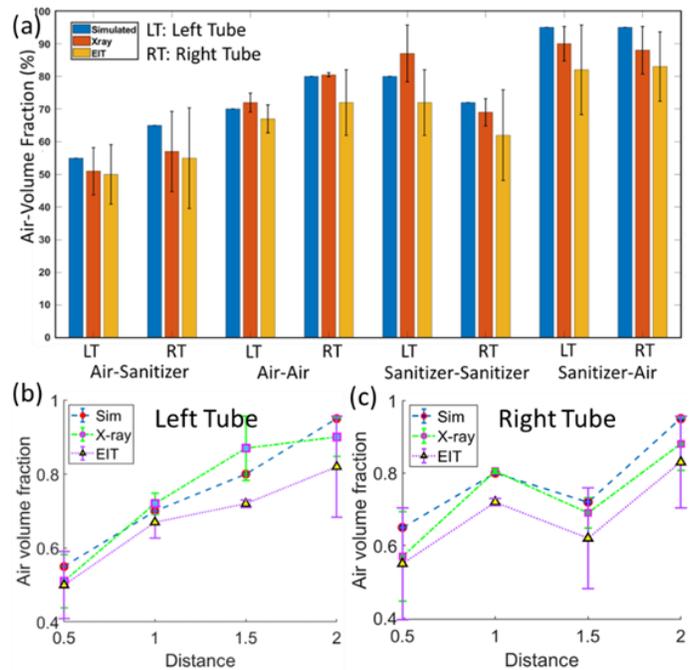

Figure 5: Air volume fraction from simulation, X-ray CT, and EIT.

## 4. Conclusion

This study seeks to provide a comprehensive understanding of (1) the design and development of in- situ multiphase flow chamber, (2) the logically correct method to set initial conditions of simulation, (3) furnish alternate modality EIT than costly slow X-ray CT with minor sacrifice of accuracy and (4) common cross-sectional fluid imaging methodology using multi-modal tomography systems. A solution is presented by using X-ray CT to (a) reconstruct the real internal boundaries of the bent/fold channel and (b) track the transient state of the multi-phase fluid profile that is later used by a simulation tool for accurate CFD analysis.

A multimodal imaging system mounted on the same cross-section capturing data simultaneously from EIT and X-ray CT is performed to estimate attenuation and impedance profile at the same time. The compact adaptor can only provide such a study as many laboratory-based X-ray CT chambers have limitations on sample or ROI size.

Here are pointwise statements that evolved during the experiments using an in-situ adaptor:

1. The proposed innovative in-situ flow adaptor may provide a cost-effective solution, offering portability and convenience for conducting multiphase flow analyses within a laboratory setting.
2. X-ray CT seems to be the gold standard, but it is a slow option. The CFD tools need further modeling advancement for accurate cross-sectional analysis, as shown by non-invasive imaging.
3. EIT has agreement with X-ray CT for common cross-sectional qualitative and quantitative analysis.
4. As far as air volume fraction deduction is concerned, the Ansys tool with suggested X-ray-assisted methodology to fix initial conditions has good agreement with EIT and X-ray, both and can be used where these cannot be used.
5. EIT can be used to differentiate conductive and non-conductive phases of the fluid, and X-ray CT can be used for identifying low to heavy mass attenuating phases of the fluid simultaneously.

## Declarations

Author Contributions: VS: EIT experiments and related data analysis, Ar: Software, Instrumentation, SS: X-ray data acquisition and visualization; MG: methodology, investigation, writing, funding. All authors have read this manuscript.

Funding: Science and Engineering Research Board (IN), Early Career Research Grant code ECR/2017/001432.

Data Availability Statement: The datasets created and/or analyzed during the current study are made available via respective weblinks and added in text.

Acknowledgments: Ms. Vaishali Sharma is thankful to CSIR, India, for awarding the SRF Fellowship.

Conflicts of Interest: "The authors declare no conflict of interest."

Dataset Location: https://dx.doi.org/10.21227/w560-8314

# References:

[1]     Multiphase Flow Handbook (Google eBook), (2005) 1156.
        https://books.google.com/books/about/Multiphase_Flow_Handbook.html?id=M0MrBgAAQBAJ.

[2]     P.Z. Júnior, G. Zanon, S. José, D. Campos, B. Alexandre, G. Garmbis, L.B. Alkmin, M.R. Richter, E.V.
        Oazen, P.N. Chaves, E. Hippert Júnior, CHALLENGES ABOUT TESTING, WELDING AND NDT OF
        CRA PIPELINES IN BRAZILIAN PRE-SALT, 2012. http://www.asme.org/about-asme/terms-of-use.

[3]     J. Valls Miro, N. Ulapane, L. Shi, D. Hunt, M. Behrens, Robotic pipeline wall thickness evaluation for dense
        nondestructive testing inspection, J Field Robot 35 (2018) 1293–1310. https://doi.org/10.1002/rob.21828.

[4]     F. Viana, Challenges of Multiphase Flow Metering in Heavy Oil Applications, 2013.

[5]     D. Bestion, The difficult challenge of a two-phase CFD modelling for all flow regimes, Nuclear Engineering
        and Design 279 (2014) 116–125. https://doi.org/10.1016/j.nucengdes.2014.04.006.

[6]     S. Subramaniam, Lagrangian-Eulerian methods for multiphase flows, Prog Energy Combust Sci 39 (2013)
        215–245. https://doi.org/10.1016/j.pecs.2012.10.003.

[7]     E. Elias, Y. Ben Haim, DETERMINATION OF SPATIAL DISTRIBUTION IN TWO-PHASE FLOW
        USING SCATTERED GAMMA RADIATION, 1980.

[8]     T. Dyakowski, Process tomography applied to multi-phase flow measurement, 1996.
        http://iopscience.iop.org/0957-0233/7/3/015.

[9]     D.H. Rothman, S. Zaleski, Lattice-gas models of phase separation: interfaces, phase transitions, and
        multiphase flow, 1994.

[10]    S. Subramaniam, Multiphase flows: Rich physics, challenging theory, and big simulations, Phys Rev Fluids
        5 (2020). https://doi.org/10.1103/PhysRevFluids.5.110520.

[11]    A. Mahvash, A. Ross, Two-phase flow pattern identification using continuous hidden Markov model,
        International Journal of Multiphase Flow 34 (2008) 303–311.
        https://doi.org/10.1016/j.ijmultiphaseflow.2007.08.006.

[12]    A. Mahvash, A. Ross, Application of CHMMs to two-phase flow pattern identification, Eng Appl Artif
        Intell 21 (2008) 1144–1152. https://doi.org/10.1016/j.engappai.2008.02.005.

[13]    T.Y. Chalfi, S.M. Ghiaasiaan, Pressure drop caused by flow area changes in capillaries under low flow
        conditions, International Journal of Multiphase Flow 34 (2008) 2–12.
        https://doi.org/10.1016/j.ijmultiphaseflow.2007.09.004.

[14]    H.S.H. Ai-Najjar, N.B. Abu, A.- Soof, Alternative Flow-Pattern Maps Can Improve Pressure-Drop
        Calculations of the Aziz et al. Multiphase-Flow Correlation, 1989.

[15]    H.M. Prasser, U. Hampel, P. Schütz, TOPFLOW pressure chamber – Versatile techniques to simplify design
        and instrumentation of thermal fluid dynamic experiments at high pressure, Nuclear Engineering and Design
        372 (2021). https://doi.org/10.1016/j.nucengdes.2020.110971.

[16]    M. Wagner, A. Bieberle, M. Bieberle, U. Hampel, Dynamic bias error correction in gamma-ray computed
        tomography, Flow Measurement and Instrumentation 53 (2017) 141–146.
        https://doi.org/10.1016/j.flowmeasinst.2016.10.012.

[17]    F. Barthel, M. Bieberle, D. Hoppe, M. Banowski, U. Hampel, Velocity measurement for two-phase flows
        based on ultrafast X-ray tomography, Flow Measurement and Instrumentation 46 (2015) 196–203.
        https://doi.org/10.1016/j.flowmeasinst.2015.06.006.

[18]    A. Ferrari, J. Jimenez-Martinez, T. Le Borgne, Y. Méheust, I. Lunati, Challenges in modeling unstable two-
        phase flow experiments in porous micromodels, Water Resour Res 51 (2015) 1381–1400.
        https://doi.org/10.1002/2014WR016384.

[19]    S. Acharya, B.R. Baliga, K. Karki, J.Y. Murthy, C. Prakash, S.P. Vanka, Pressure-based finite-volume
        methods in computational fluid dynamics, J Heat Transfer 129 (2007) 407–424.
        https://doi.org/10.1115/1.2716419.

[20]    W. Jeong, J. Seong, Comparison of effects on technical variances of computational fluid dynamics (CFD)
        software based on finite element and finite volume methods, Int J Mech Sci 78 (2014) 19–26.
        https://doi.org/10.1016/j.ijmecsci.2013.10.017.

[21]    E. Bellenger, P. Coorevits, Controlled cost of adaptive mesh refinement in practical 3D finite element
        analysis, Advances in Engineering Software 38 (2007) 846–859.
        https://doi.org/10.1016/j.advengsoft.2006.08.035.

[22]    J. Lantz, V. Gupta, L. Henriksson, M. Karlsson, A. Persson, C.J. Carlhäll, T. Ebbers, Intracardiac flow at 4D
        CT: Comparison with 4D flow MRI, Radiology 289 (2018) 51–58, doi:10.1148/radiol.2018173017.

[23]    L.T. Jorgensen, M.S. Traberg, M.B. Stuart, J.A. Jensen, Performance Assessment of Row-Column

Transverse Oscillation Tensor Velocity Imaging Using Computational Fluid Dynamics Simulation of Carotid Bifurcation Flow, IEEE Trans Ultrason Ferroelectr Freq Control 69 (2022) 1230–1242. https://doi.org/10.1109/TUFFC.2022.3150106.

[24] G. Pendharkar, R. Deshmukh, R. Patrikar, Investigation of surface roughness effects on fluid flow in passive micromixer, Microsystem Technologies 20 (2014) 2261–2269. https://doi.org/10.1007/s00542-013-1957-y.

[25] T. Zhang, G.N. Barakos, Assessment of implicit adaptive mesh-free CFD modelling, Int J Numer Methods Fluids 96 (2024) 670–700. https://doi.org/10.1002/FLD.5266.

[26] N.T. Clemens, FLOW IMAGING, n.d.

[27] R. Huang, W. Liu, J. Cheng, J. Wu, Measurement of hypersonic turbulent boundary layer on a flat plate using cylindrical focused laser differential interferometer, Physics of Fluids 35 (2023). https://doi.org/10.1063/5.0141681.

[28] J.J. van Wijk, Hanwei. Shen, S.C. North, Institute of Electrical and Electronics Engineers., SIGGRAPH., IEEE Computer Society. Technical Committee on Visualization and Graphics., European Association for Computer Graphics., IEEE PacificVis 2010 : IEEE Pacific Visualization Symposium 2010, Taipei, Taiwan, March 2-5, 2010 : proceedings, IEEE, 2010.

[29] J. Sheng, H. Meng, R.O. Fox, Validation of CFD simulations of a stirred tank using particle image velocimetry data, Canadian Journal of Chemical Engineering 76 (1998) 611–625. https://doi.org/10.1002/cjce.5450760333.

[30] E.A. Afolabi, J.G.M. Lee, Investigating the Effect of Air Volume Fraction on the Velocity Distributions of Air-Water flow in a Pipe Separator, 2013. http://www.ijsat.com36.

[31] S.Z. Islami Rad, R. Gholipour Peyvandi, S. Sadrzadeh, Determination of the volume fraction in (water-gasoil-air) multiphase flows using a simple and low-cost technique: Artificial neural networks, Physics of Fluids 31 (2019). https://doi.org/10.1063/1.5109698.

[32] A. Lopez, R. Bacelar, I. Pires, T.G. Santos, J.P. Sousa, L. Quintino, Non-destructive testing application of radiography and ultrasound for wire and arc additive manufacturing, Addit Manuf 21 (2018) 298–306. https://doi.org/10.1016/j.addma.2018.03.020.

[33] E.A. Zwanenburg, M.A. Williams, J.M. Warnett, Review of high-speed imaging with lab-based x-ray computed tomography, Meas Sci Technol 33 (2022). https://doi.org/10.1088/1361-6501/ac354a.

[34] T.J. Heindel, A review of X-ray flow visualization with applications to multiphase flows, Journal of Fluids Engineering, Transactions of the ASME 133 (2011). https://doi.org/10.1115/1.4004367.

[35] A. Adler, R. Guardo, Electrical Impedance Tomography: Regularized Imaging and Contrast Detection, 1996.

[36] S. Mansouri, Y. Alharbi, F. Haddad, S. Chabcoub, A. Alshrouf, A.A. Abd-Elghany, Electrical Impedance tomography – Recent applications and developments, J Electr Bioimpedance 12 (2021) 50–62. https://doi.org/10.2478/joeb-2021-0007.

[37] D.L. George, A Review of Electrical Impedance Techniques for the Measurement of Multiphase Flows, 1996. http://fluidsengineering.asmedigitalcollection.asme.org/.

[38] R. Babaei, B. Bonakdarpour, F. Ein-Mozaffari, The use of electrical resistance tomography for the characterization of gas holdup inside a bubble column bioreactor containing activated sludge, Chemical Engineering Journal 268 (2015) 260–290. https://doi.org/10.1016/j.cej.2015.01.042.

[39] R.E. Vieira, M. Parsi, B.S. McLaury, S.A. Shirazi, C.F. Torres, E. Schleicher, U. Hampel, Experimental characterization of vertical downward two-phase annular flows using Wire-Mesh Sensor, Chem Eng Sci 134 (2015) 324–339. https://doi.org/10.1016/j.ces.2015.05.013.

[40] A. Bieberle, H.U. Härting, S. Rabha, M. Schubert, U. Hampel, Gamma-ray computed tomography for imaging of multiphase flows, Chem Ing Tech 85 (2013) 1002–1011. https://doi.org/10.1002/cite.201200250.

[41] S.H. Stavland, Y. Arellano, A. Hunt, R. Maad, B.T. Hjertaker, Multimodal Two-Phase Flow Measurement Using Dual Plane ECT and GRT, IEEE Trans Instrum Meas 70 (2021). https://doi.org/10.1109/TIM.2020.3034615.

[42] Arduino Bluetooth Control - Apps on Google Play, (n.d.). https://play.google.com/store/apps/details?id=com.broxcode.arduinobluetoothfree&hl=en&gl=US.

[43] O. Luppi Silva, R. Gonzalez Lima, T. Castro Martins, F. Silva de Moura, R. Seiji Tavares, M. Sales Guerra Tsuzuki, Influence of current injection pattern and electric potential measurement strategies in electrical impedance tomography, Control Eng Pract 58 (2017) 276–286. https://doi.org/10.1016/j.conengprac.2016.03.003.

[44] S. Orvalho, M.C. Ruzicka, G. Olivieri, A. Marzocchella, Bubble coalescence: Effect of bubble approach velocity and liquid viscosity, Chem Eng Sci 134 (2015) 205–216. https://doi.org/10.1016/j.ces.2015.04.053.